
\magnification=1200
\input amssym.def
\input amssym
\def\ra{\rightarrow}
\def\Rr{\Bbb R}

\def\Zz{\Bbb Z}
\def\Pp{\Bbb P}

\hfill FM 90/92
\vskip3truecm
\centerline{\bf TOPOLOGICAL  ``OBSERVABLES"}

\centerline{\bf IN}

\centerline {\bf SEMICLASSICAL FIELD THEORIES}

\vskip 1.5truecm

\centerline{\bf Margherita Nolasco and Cesare Reina}
\vskip 0.3cm
\centerline { S.I.S.S.A. - {\it International School for Advanced Studies}}

\centerline{Strada Costiera 11 - TRIESTE (Italy)}

\vskip 3truecm
{\bf Abstract.} We give a geometrical set up for the semiclassical
approximation to euclidean field theories having families of minima
(instantons) parametrized by suitable moduli spaces ${\cal M}$.
The standard examples are of course Yang-Mills theory and non-linear
$\sigma$-models. The relevant space here is a family of measure spaces
$\tilde {\cal N} \ra {\cal M}$, with standard fibre a distribution
space, given by a suitable extension of the normal bundle to ${\cal M}$
in the space of smooth fields. Over $\tilde {\cal N}$ there is a probability
measure $d\mu$ given by the twisted product of the (normalized)
volume element on ${\cal M}$ and the family of gaussian measures with
covariance given by the tree propagator $C_\phi$ in the background of an
instanton $\phi \in {\cal M}$. The space of ``observables", i.e. measurable
functions on ($\tilde {\cal N}, \, d\mu$), is studied and it is shown to
contain a topological sector, corresponding to the intersection theory on
${\cal M}$. The expectation value of these topological ``observables"
does not depend on the covariance; it is therefore exact at all
orders in perturbation theory and can moreover be computed
in the topological regime by setting the covariance to zero.

\vfill

\centerline{E-mail: NOLASCO@ITSSISSA.BITNET,
REINA@ITSSISSA.BITNET}
\eject

\noindent {\bf 1. Introduction.}
\vskip3truemm

The basic idea of quantizing a field theory in presence of instantons goes
back to `t Hooft [`tH] in his pioneering papers on Yang Mills theory. It
includes from the very beginning some non linearity in the space of fields,
represented by the instanton moduli space ${\cal M}$.
More recently topological field theories [W] were studied as an attempt to
give a field theoretical meaning to the intersection theory on certain moduli
spaces.

In this paper we will show that in fact topological ``observables" arise at the
semiclassical level also in non topological field theories, provided the
minima of the classical action occour in families. Our set up is a
generalization of the result of Labastida [L] and at the same time it
gives a full mathematical status to the path integrals entering the game.
At the semiclassical level this can be done in a quite standard way in terms
of gaussian measures, for which we refer e.g. to Gel'fand and Vilenkin's book
[GV]. The non linearity entailed by the instanton moduli space
${\cal M}$ gives rise to a non gaussian measure $d\mu$ on the dual
$\tilde {\cal N}$ of the normal bundle to ${\cal M}$ in the space of smooth
fields. Roughly speaking $d\mu$ is the ``product" of a family $d\mu_{\phi}$
of gaussian measures on the fibres $\tilde N_{\phi}$ of $\tilde {\cal N}$
times a volume element $dv$ on ${\cal M}$ which represents the non linear
contribution to the measure.
Section 2 is devoted to the geometric construction of the
measure $d\mu$. Concrete examples will be discussed in sect 3.

Although $\tilde {\cal N}$ has to be considered as a family
$\bigcup _{\phi \in {\cal M}} \tilde N_{\phi} $ of a measure spaces, we
will think of it as if it was a vector bundle over ${\cal M}$. With this
analogy in mind, it turns out that the family ${d\mu}_{\phi}$ of gaussian
measures plays the ${\rm r\hat ole}$ of the Thom class of $\tilde {\cal N}$. To
make
this concrete, we will introduce in section 4 a cohomology explicitly
designed for the job.
{}From the physical point of view, the coboundary operator $\tilde d$ of this
cohomology is the functional analogue of the BRS operator. Moreover,
an analogue of Thom
isomorphism goes through, identifying the
BRS cohomology of $\tilde d$ with the
cohomology of the moduli space ${\cal M}$.
In particular, to any cycle on ${\cal M}$ one can
associate an observable and to the expectation value
of products of such obvervable equals the intersection number
of the corresponding cycles, whenever defined.

\vskip2truecm
\noindent {\bf 2. Geometric set up }
\vskip3truemm

The basic ingredients to set up a ``euclidean" classical field theory
are

\noindent
i) a space of classical field $\Phi = \{ \phi : X \to Y \}$ defined
on a compact manifold X without boundary with values in a manifold Y.
X and Y are assumed to have
further structures, as Riemannian metrics etc., such that one can give to
$\Phi$ the structure of a Hilbert manifold. In particular the
tangent space $T_{\phi}\Phi$ will be assumed to be a Hilbert space. In all the
physical examples it can be identified with the Sobolev
space of section of a suitable vector bundle, $E_{\phi}$ say, over X which are
square integrable together with their derivatives up to order s. We will denote
by $< \, , \, >_{s}$ the inner product of Sobolev index s. Since $X$
is compact without boundary, the family of inner products $< \, ,\,
>_s$ for $s\geq 0$ gives $T_{\phi}\Phi$ the structure of a nuclear
space. This will be crucial in order to construct the gaussian measures
we will need below.

\noindent
ii) an action functional $S: \Phi \to {\Rr }$. When this is $C^{2}$ on $\Phi$,
one can write the first as well as the second variation of $S$ as
$$
dS(\delta\phi)|_\phi =<EL_{\phi },\delta\phi>_0
$$
$$
QS(\delta\phi ,\delta\phi)|_\phi =<\delta\phi,H_{\phi }\delta \phi>_0,
$$
where $EL_{\phi}$, $H_{\phi}$ are
the Euler--Lagrange expression and the Hessian operator of $S$
at $\phi \in \Phi$.

We are interested in studying the cases in which $H_{\phi }$
has kernel. In all physical examples $H_{\phi}$ turns out to be selfadjoint
elliptic operator ( may be modulo the action of a group ${\cal G}$ as for gauge
theories, in which cases $\Phi$ will be the space of orbits of the group
${\cal G}$, see sect. 3.1 for more details). We have then an obvious exact
sequence
$$
0\ra kerH_{\phi} \ra T_{\phi}\Phi \ra N_{\phi} \ra 0,
$$
which gives the space $N_{\phi}$ of the classical variations
of $\phi$ along which the Hessian is invertible.
We say that $\phi$ is a minimum (an instanton) for $S$ if $EL_{\phi}=0$ and
$H_{\phi}$ is positive definite on $N_\phi$.
Any element $\xi \in ker H_{\phi}$ is a Jacobi field along $\phi$,
i.e. it is such that $\phi + \epsilon \xi$ is still a minimum (at the first
order in $\epsilon$). We will denote by ${\cal M} \subset \Phi$ a connected
component of the moduli space of instantons. This will be assumed to be a
manifold, i.e. we work on the subset of instantons which can be given a
manifold structure.
Then ${\rm ker} \, H_{\phi}$ can be identified with the tangent space
$T_{\phi}{\cal M}$ and $N_{\phi}$ is
the fibre at $\phi$ of the normal bundle ${\cal N}$ w.r.t. the embedding
${\cal M} \ra \Phi$ of the moduli space in the space of all classical fields.

To construct the semiclassical approximation to the quantized version of the
theory, we need first to give meaning to formal path integrals as
$$
" \,  \int D\eta \, e^{-<\eta ,H_\phi \eta>} . "
$$
This can be done by introducing well defined gaussian measures; i.e.
by defining the covariance operator (i.e. the ``euclidean
propagator") on the space $N_{\phi}^{'}$ dual to $N_{\phi}$. $N_{\phi}^{'}$
will play the ${\rm r\hat ole}$ of the space of the currents (i.e. test
function)
which we want to keep as smooth as possible.
Let $E^*_{\phi}$ be the bundle on $X$ dual to $E_\phi$; any
$\eta \in T_{\phi}\Phi = C^{\infty}(X,E_{\phi})$ defines a functional on
$C^{\infty}(X,E_{\phi}^{*})$ by setting
$F_{\eta}(j) = \int j\cdot \eta * {\bf 1}$; $* {\bf 1}$
being a fixed volume element on X. Accordingly we will identify
$C^{\infty}(X,E^{*}_{\phi})$ with the cotangent space $T_{\phi}^{*}\Phi$.
The Hessian gives us a transposed operator $H_{\phi}^{'}$ such that
$F_{H_{\phi}\eta}(j)=F_{\eta}(H^{'}_{\phi}j)$.
Again we have an exact sequence
$$
0 \ra {\rm ker} \, H^{'}_{\phi} \ra T^*_{\phi}\Phi \ra N'_{\phi} \ra 0,
$$
which identifies ${\rm ker} \, H^{'}_{\phi}$ with the cotangent space
$T^{*}_{\phi}{\cal M}$ and $N'_{\phi}$ with the fibre at $\phi$ of the
conormal bundle to ${\cal M}$ in $\Phi$. As $E_{\phi}$ is assumed to have
smooth fibre metric $h_{\phi}$ we have an isomorphism
$C^{\infty}(X,E_{\phi}^{*}) \simeq C^{\infty}(X,E_{\phi})$ given by
$ j\mapsto h_{\phi}(j, \cdot )$. Under such an isomorphism
$H^{'}_{\phi}$ transforms into the adjoint $H^*_{\phi}=H_{\phi}$ so
${\rm Ker} H^{'}_{\phi} \simeq {\rm ker} H_{\phi}^*$ is actually the
isomorphism $T_{\phi}^{*}{\cal M} \simeq T_{\phi}{\cal M}$.

The $L^{2}$ inner product $< \, ,\, >_{0}$ on $C^{\infty}(X,E_{\phi})$ given by
$$
<\eta_{1},\eta_{2}>_0 = \int_{X} h_{\phi}^{-1}(\eta_{1},\eta_{2}) *{\bf 1}
$$
induces a metric g on ${\cal M}$, called the Weil--Peterson metric, which we
assume to be smooth. Thus we have a smooth volume element $*_{g}{\bf 1}$
on ${\cal M}$. We will also assume, as it is the case in all concrete
examples, that ${\cal M}$ has
a natural compactification $\bar {\cal  M}$  and that
$\int_{\bar {\cal M}} *_g {\bf 1} = {\rm vol}(\bar {\cal M})$ is finite.
Thus, with the normalized measure $dv =*_g {\bf 1}/{\rm vol}(\bar {\cal M})$,
$\bar {\cal M}$ itself is assumed to be a probability space.

On the fibres $N'_{\phi}$ the transposed Hessian $H'_{\phi}$ can be inverted,
giving a positive definite operator $C_{\phi}=H_{\phi}^{'-1}$ and a positive
quadratic form $<j,C_{\phi}j>$. Now $\forall \phi \in {\cal M}$,  $N'_{\phi}$
is a nuclear space and the functional $W_{\phi}(j) = e^{-<j,C_{\phi}j>}$ is
continuous, positive and $W_{\phi}(0)=1$. Accordingly, it is the
Fourier transform of a positive countably additive measure on the topological
dual $\tilde N_{\phi}$ of $N_{\phi}$, i.e.
$$
W_{\phi}(j)= \int d\mu_{\phi} \, e^{i\eta (j)}.
$$
The measure $\mu_{\phi}$ is called the gaussian measure with covariance
$C_{\phi}$. Recall that generically $\tilde N_{\phi}$ will be a distribution
space given by the union of the completions $N_{\phi ,s}$ of
$N_{\phi}$ w.r.t. the Sobolev norm $< \, , \,>_{s}$ with $s\in\Zz$.
Summing up we have the following :

\vskip3truemm
\noindent {\bf Proposition.} Let $S:\Phi \ra \Rr$ be
a classical action, with a selfadjoint elliptic non negative Hessian operator
$H_{\phi}$ along a family of instantons parametrized by
a smooth manifold ${\cal M}$. Then there is a family of gaussian
measures $d\mu_{\phi}$ on the topological dual  $\tilde N_{\phi}$ of the
fibres $N_{\phi}$ of the
nuclear completion of the conormal bundle to ${\cal M}$ in $\Phi$.
\vskip3truemm

As the measure $d\mu_{\phi}$ is automatically normalized,
we have a family of probability spaces, parametrized by
$\phi \in {\cal M}$. We recall that each of them is a triple
 ($\tilde N_{\phi},\Sigma_{\phi},\mu_{\phi}$), where $\Sigma_{\phi}$
is the smallest $\sigma$-algebra generated by the cylinder
sets [Y]. The normal ``bundle"
$\tilde {\cal N}=\cup _{\phi \in {\cal M}} \tilde N_{\phi}$
has a natural probability measure $d\mu$, given by the family $d\mu_{\phi}$
and the (normalized) volume element $dv$ on ${\cal M}$. This is constructed as
follows. A subset $U\subset \tilde {\cal N}$ is measurable if $\pi(U)$ is a
Borel set in ${\cal M}$ and if $U_{\phi}=U \cap \tilde N_{\phi}$ is measurable
with measure $f(\phi )=:\mu_{\phi}(U_{\phi})\in {\cal L}^1({\cal M},dv)$.
We will denote by $\Sigma$ the smallest $\sigma$-algebra generated by such
sets. Then by definition the measure $d\mu$ is such that
$$
\int _U d\mu = \int _{\pi(U)} dv f(\phi ).
$$
The triple $(\tilde N, \Sigma, \mu)$ is a probability space.

Notice that formally
$$
\int_{\tilde {\cal N}} ... d\mu = \int_{\cal M} dv \int _{\tilde N_{\phi}}
D\eta ... e^{-<\eta ,H(\phi )\eta >} \, .
$$
As is well know, neither the "Lebesgue measure" $D\eta$ nor the
quadratic form $<\eta ,H_{\phi} \eta>$ exist separately on $\tilde N_{\phi}$,
nevertheless $d\mu$ is well defined.
The physical consequence of such a construction is that,
whenever all the assumptions made above hold true,
the semiclassical measure $d\mu$ is well defined and gives a mathematical
meaning to our path integrals.

\vfill\eject
\noindent {\bf 3. Examples}
\vskip5truemm

In this section we briefly discuss some examples of the geometric set up
given above.
\vskip 0.5 truecm
\noindent 3.1 -- As mentioned in the introduction, the standard example is
euclidean gauge theory on an SU(2)--principal bundle P over a compact
Riemannian 4--manifold X.
We will take as space of classical fileds the space of gauge orbits
$\Phi= \bar {\cal C}/\bar {\cal G}$, i.e.
the space $\bar {\cal C}$ of irreducible connections on P modulo the
action
of the quotient $\bar {\cal G} ={\cal G}/{\cal Z}$ of the group ${\cal G}$ of
gauge transformations by its centre ${\cal Z}$. We refer to Singer [S],
Donaldson and Kronheimer [DK] or to Grossier and Parker [GP] for details.
The tangent bundle $T\Phi=\bar {\cal C} \times_{\bar {\cal G}}
Ker\, d_{A}^{*}$ has the typical fibre at the orbit through $\phi=[A]$
isomorphic to $Ker\, d_{A}^{*} \subset \Omega^{1}(ad \, P)$. Here
$\Omega^p(adP)$ denotes the space of smooth p-forms on $X$ with values in
the bundle $adP$ associated to $P$ via the adjoint action of its
structure group on its Lie algebra.
The Yang--Mills action $S: \Phi \to
{\Rr }$, given by $S(\phi)= \Vert F_{A} \Vert^{2}$, has first and second
variations
$dS(\delta \phi)=<d_{A}^{*} F, \delta \phi>,
QS(\delta \phi,\delta \phi) = <\delta \phi, L_{A} \delta \phi>$ with
Hessian
$$L_{A}=d_{A}^{*}d_{A} + \hat F $$
(see e.g. Atiyah--Bott [AB]). As is well known, $L_A$ is elliptic
modulo the action of the gauge group, i.e. it becomes elliptic on the
gauge fixing plane $kerd_A^*\subset T_A{\cal C}$. Indeed on this plane
$L_A=H_A=:d_A^*d_A+d_Ad_A^*+\hat F$, which is obviusly elliptic. Hence
is kernel is finite dimensional and there is an
isomorphism $kerH_A\simeq T_{[A]}{\cal M}$.

Let $\tilde {\cal M}$ be a component of the space
of all (irreducible) instantons in
$\bar {\cal C}$. As is well known, it is a principal $\bar {\cal
G}$-bundle over the moduli space ${\cal M}$. At any $A\in \tilde {\cal
M}$ we have a $\bar {\cal G}$-equivariant
decomposition $\Omega ^1(adP)= kerH_A\oplus N_A\oplus
Imd_A$ into orthogonal closed subspaces. Here $Imd_A$ is the tangent
space to the gauge orbit through $A$, and $N_A$ is isomorphic
to the fibre of the normal
bundle $N_{[A]}$ at $[A]$ with respect to the embedding of the
moduli space ${\cal M}$ in the space of gauge orbits $\Phi$.
Accordingly, the vector bundle $V=\tilde {\cal M}\times
_{\bar {\cal G}}\Omega^1(ad P)$ on ${\cal M}$ decomposes as
$V\simeq T{\cal M}\oplus {\cal N}\oplus {\cal I}$, where ${\cal I}= \tilde
{\cal M} \times_{\bar{\cal G}} \Omega^{0}(ad P)$.

Now the gauge fixed Hessian $H_A$ is invertible on the much
larger space $T_A{\cal C}/kerH_A\simeq N_A\oplus Imd_A$,
and
$$H_A= \matrix{L_A        &on\; N_A\cr
               d_Ad_A^*   &on\; Imd_A.\cr}$$
Decomposing $\eta\in N_A\oplus Imd_A$ as $\eta =\eta_0+\eta_1$,
the quadratic form associated to $H_A$ reads
$$<\eta,H_A\eta >=<\eta _0,H_A\eta _0>+<d_A^*\eta_1,d_A^*\eta_1>$$
and is obviously positive on this domain.
Being A irreducible, $Imd_A$ is isomorphic to the Lie algebra
$\Omega ^0(adP)$ of the group of gauge transformations, and "changing
variables" from $Imd_A$ to $\Omega^0(adP)$ is the basis of the
Faddeev-Popov procedure. Indeed, if we set $\eta_1=d_A\xi$, $\xi \in
\Omega^0(adP)$, the quadratic form above becomes
$$<\eta,H_A\eta >=<\eta_0,H_A\eta_0>+<\xi ,\vartriangle_A ^2\xi >$$
on $N_A\oplus \Omega^0(adP)$, the operator $\vartriangle_A =d_A^*d_A$
being the laplacian on $\Omega^0(adP)$.

Now the insertion of the Faddeev-Popov determinant in the formal path
integral for gauge theories has the effect of "normalizing" the formal
gaussian measure on the $\xi $-variables; indeed
$$"det\vartriangle_A \int D\xi e^{-<\xi ,\vartriangle_A ^2\xi>}=1".$$
The analogue of this procedure in our set up is then to consider the
complete gaussian measure associated to the operator $H_A\oplus
\vartriangle_A^2$ on $N_A\oplus \Omega^0(adP)$.
The transposed of this operator  gives rise to a covariance
which is the direct sum of the covariance induced by $H'_A$
on $N'_A$ plus that induced by $\vartriangle_A^2$ on $\Omega^0(adP)'$.
The associated generating functionals $W_A(j)$ induce
gaussian measures on the topological duals $N_A$ and
$\Omega^0(adP)$ and the product measure on their direct sum.
Since $W_A(j)$ is invariant under the action of the group of
gauge transformations, the induced measures are invariant and
descend to a family of measures on $\tilde{\cal N} \oplus \tilde{\cal I}$.

\vskip 0.5 truecm

\noindent 3.2 - Next we briefly mention another example, namely
semiclassical gravity around Einstein metrics. The space of classical
fields is the space of metrics $g$ with unit volume and with no
conformal automorphisms on a compact closed $n$-dimensional manifold
$(n\geq 3)$. On this space,
Einstein metrics are critical points of
the Hilbert-Einstein action
$$S(g)=\int R_g *_g1,$$
where $R_g$ is the scalar curvature and $*_g1$ is the volume element of
$g$. As is well known [Be],  the second variation of this functional
is the quadratic form associated to an Hessian operator $H_g$
which is elliptic
modulo the action of the group of diffeomorphisms. The standard gauge
fixing procedure  (which amounts to restricting the space of variations
of the metrics to the orthogonal complement of the image of the
traceless part of their Lie derivatives, see e.g. [E]) yields an
elliptic operator whose kernel is isomorphic to the tangent space
to the moduli space of Einstein metrics.
Although there are scalings of the metric along which the action $S(g)$
decreases, restricting on the space of metrics with unit volumes
gets rid of such modes, makes the Einstein metrics minima for the
action and hence yields a  positive semidefinite Hessian.
So, there is a strict analogy with the previous example. As a word
of caution, notice that in this case the space of all fields is not
an affine space and that the gauge fixing is more complicated due
to the constant volume condition; these facts may imply
subtelties wich deserve a closer study.

\vskip 0.5 truecm
\noindent 3.3 -- We will spend some more words for the case of
$\sigma$--models, i.e. harmonic theory for maps $\phi : C \to Y$ of an
algebraic curve C with values in a compact
K$ \ddot {\rm a}$hler manifold Y. To be
concrete we will stick to the case of $\phi : {\Pp }^{1} \to {\Pp }^{n}$,
${\Pp}^{n}$ being the n--dimensional complex projective space. We refer to
Palais [P] for the construction of the manifold $\Phi$ of such maps of a given
degree d, and we
limit ourselves to notice that the local model for $\Phi$ around $\phi$, i.e.
the tangent space $T_{\phi}\Phi$ is actually isomorphic to
$C^{\infty}(\phi^{*} T{\Pp }^{n})$, $T{\Pp }^{n}$ being the real tangent
bundle
to ${\Pp }^{n}$. The Dirichelet functional $S: \Phi \to {\Rr }$ reads
$S(\phi)= \int_{{\Pp }^{1}} {\rm tr} \, \alpha_{\phi}$, where $\alpha_{\phi}$
is the first fundamental form of $\phi$ w.r.t. the Fubini--Study metric on
${\Pp }^{n}$, and tr denotes the trace w.r.t. the Fubini--Study metric on
 ${\Pp }^{1}$ (or any other metric conformally equivalent to it). The
explicit form of the first
and the second variations of S are given by e.g.
Siu and Yau [SY] Eells and Wood [EW]).

Thanks to the complex structures of both the source and the target space, one
easily sees that $S(\phi) \geq {\rm deg }\, \phi =\int_{{\Pp }^{1}} \phi^{*}
\omega$, $\omega$ being the Fubini-- Study  K$\ddot {\rm a}$lher form, and that
the equality holds precisely if $\phi$ is holomorphic (or antiholomorphic).
The holomorphic maps are called instantons, and let ${\cal M}$ denote the set
of such maps of a fixed degree {\it d}. Since $S(\phi)=d$ for any
$\phi \in {\cal M}$ and $S(\phi) > d $ for $\phi \in \Phi - {\cal M}$, we see
that the Hessian has as kernel at $\phi$ the space of holomorphic deformations
of $\phi$, i.e. ${\rm ker} \, H_{\phi} \simeq H^{0}({\Pp }^{1},
\phi^{*} T'{\Pp }^{n})$ (where $T'\Pp ^n$ is the holomorphic tangent
bundle) and we have an exact sequence
$$0 \ra H^{0}({\Pp }^{1}, \phi^{*} T'{\Pp }^{n}) \ra
C^{\infty}({\Pp }^{1}, \phi^{*} T'{\Pp }^{n}) \ra N_{\phi} \ra 0
$$
where $H^{0}({\Pp }^{1}, \phi^{*} T{\Pp }^{n})$ and $N_{\phi}$ play the
r$\hat o$le of the tangent space $T_{\phi}{\cal M}$ and the normal space to
${\cal M}$ in $\Phi$, assuming of course that ${\cal M}$ is a manifold.
This is indeed the case, as shown for instance by [St] by applying the
techniques of {\it Quot} schemes of Groethendieck. If we think of
${\Pp }^{n}$ as
the dual projective space, ${\cal M}$ is actually isomorphic to a projective
space itself, i.e. ${\cal M} \simeq {\Pp }^{(n+1)d + n }$.

These results actually hold true for more general Grassmannian--valued
$\sigma$--models (see again [St] for more details). The cohomology ring of such
${\cal M}$'s is also known, being a quotient of the cohomology
ring of a product of two Grassmannians. In view of the results of Gepner's [G],
it is tempting to imagine that this ring may be identified with some chiral
ring in superconformal field theory.

\vskip 0.5 truecm
\noindent 3.4 - A somewhat different class of examples which
fits in a similar scheme has to do with ``free" fields on an
external background. To quote a couple of these, one may consider some
bosonic fields f, e.g. $f \in C^{\infty}(X,F)$, F being an hermitean vector
bundle associated to P via some unitary representation $\rho$ and taking an
action $S_{A}(f)= \Vert d_{A}f \Vert^{2}$ in the background of an instanton A.
Or else on a Riemann surface C of genus $g \geq 2$, pick up a metric {\it m} of
constant curvature $R=-1$ modulo diffeomorphisms. If $t$ is a smooth tensor
field on C one can study the action $S_{g}(t)= \Vert d_{g}t \Vert^{2}$,
where, as {\it g} varies in the Teichm$\ddot {\rm u}$ller space,
$d_g$ is the Levi--Civita
connection. In all such
cases one has a family of distribution spaces parametrized by suitable
moduli and a gaussian measure on each of them. Of course there will be no more
normal bundles, but the ideas of this paper obviously apply.

\vskip 2truecm

\noindent {\bf 4. Gaussian cohomology and BRS operator}
\vskip5truemm

Having constructed a measure on $\tilde {\cal N}$, we next turn to
study the ``observables", which we assume to be
all the measurable functions ${\cal O}:\tilde {\cal N} \ra \Rr$,
i.e. ${\cal O}\in {\cal L}^{1}(\tilde {\cal N},d\mu)$.
Here we use quotation marks because, as we will shortly
see, such measurable functions include objects which are by no means local in
the usual sense. Although there is no perturbative clue as for their physical
meaning, they will contain  the ``topological sector" of the theory.

To see how this works, let us first notice that the family of measures
$d\mu_{\phi}$ plays the ${\rm r\hat ole}$ of the Thom class of $\tilde
{\cal N}$. To avoid confusion, in our case $\tilde {\cal N}$  is simply a
family of measure spaces, with a projection
$\pi: \tilde {\cal N} \to {\cal M}$. We will not need to give $\tilde {\cal N}$
the structure of a vector bundle, even if sometimes we will speak as if it was.
Let us introduce the space
$$
{\cal L}^{1}(\tilde {\cal N},d\mu_{\phi}) =
\left \{ W:\tilde {\cal N} \to {\Rr } \, \vert \,
\omega(\phi) = \int_{\tilde N_{\phi}}
W d\mu_{\phi} \in C^{\infty}({\cal M}) \right \}.
$$
Any object of the form $W d\mu_{\phi}$, with
$W \in {\cal L}^{1}(\tilde N,d\mu_{\phi})$ will be called a vertical "form"
on $\tilde {\cal N}$, with gaussian behavior on the fibers and vertical
"codimension" 0. This last statement means that its restriction to each fibre
$\tilde N_{\phi}$ is actually a measure on it. We will denote by
$\Omega^{0,0}(\tilde {\cal N}) \simeq
{\cal L}^{1}(\tilde {\cal N},d\mu_{\phi})$ the space of such forms.
Notice that $\Omega^{0,0}(\tilde {\cal N})$ is a $C^{\infty}({\cal M})$ module,
since $\forall f \in C^{\infty}({\cal M}),
\,\,\, \int f W d\mu_{\phi} = f \int W d\mu_{\phi}$.
Accordingly we can construct the tensor products
$$
\Omega^{p,0}(\tilde {\cal N}) = \Omega^{p}({\cal M})
\otimes_{C^{\infty}({\cal M})} \Omega^{0,0}(\tilde {\cal N}),
$$
$\Omega^{p}({\cal M})$ being the $C^{\infty}({\cal M})$--module of smooth
p--forms on ${\cal M}$. The first step in making contact with the Thom class
is to notice that there is a map
$$
\eqalign {
T:\Omega^{p}({\cal M}) &\to \Omega^{p,0}(\tilde {\cal N}) \cr
\omega &\mapsto \omega \otimes d\mu_{\phi} \cr },
$$
such that $\int_{\tilde N_{\phi}} \cdot \,\, T= {\rm id}$
on $\Omega^{p}({\cal M})$.

Next we need to introduce the space $\Omega^{p,1}(\tilde {\cal N})$ of
gaussian "forms" on $\tilde {\cal N}$ of vertical "codimension" 1.
For this we will assume that ${\cal N}'$ has trivial subline bundles $L_{i}$,
or in other words that there are global non vanishing sections $j_i$ of
${\cal N}'$. Although one might avoid such an assumption, we notice that this
is
the case for the physical examples (see appendix A), and that it helps in
making the following arguments quite simple.
Let $A_{i}(\phi) = \left \{ \eta \in \tilde N_{\phi} \vert
\eta(j_{i}) \vert_{\phi}=0 \right \}$ be the annihilator of $j_{i}$ at $\phi$.
Projecting the covariance to $N'_{\phi}/{\Rr } \{ j_{i}|_{\phi}\}$ we get by
Fourier transform a gaussian measure $d\mu_{i}$ on $A_{i}(\phi)$. Now
$\tilde N_{\phi}$ is a cylinder over the real line
$\tilde N_{\phi}/A_{i}(\phi)$ and we can consider the family of the
(unnormalized) gaussian ``measures" ${\rm e}^{-q_{i}^{2} / 2}d\mu_{i}$ on
$\tilde N_{\phi}$, parametrized by $q_{i}=\eta(j_i)$,
$\eta \in \tilde N_{\phi}/A_{i}(\phi)$.
Clearly enough $\int_{\tilde N_{\phi}} {\rm e}^{-q_{i}^{2} / 2}d\mu_{i}=0$.
We will consider objects of the form
$$
\alpha = \sum_{i=1}^{n} W_{i} {\rm e}^{-q_{i}^{2} / 2}d\mu_{i},
$$
with $n < \infty$, $W_{i} \in {\cal L}^{1}(\tilde N,d\mu_{\phi})$. Notice that,
as a function of $q_{i}$, $W_{i}$ is a linear combination of squared Hermite
polynomials, and
is therefore differentiable w.r.t. $q_{i}$. We will denote by
$\Omega^{0,1}(\tilde {\cal N})$ the space of such ``forms" of vertical
"codimension" 1.
Again we set $\Omega^{p,1}(\tilde {\cal N})= \Omega^{p}({\cal M}) \otimes
\Omega^{0,1}(\tilde {\cal N})$. We will not need to introduce ``forms" of
higher codimension.

Let us introduce the operator (for q=0,1)
$$
\tilde d : \Omega^{p,q} \to \Omega^{p+1,q} \otimes \Omega^{p,q+1}
$$
by linearly extending
$$
\eqalign {
&\tilde d (\omega \otimes W d\mu_{\phi}) = d\omega \otimes W d\mu_{\phi} +
(-)^p \omega \wedge dW d\mu_{\phi} \cr
&\tilde d (\omega \otimes W_{i}{\rm e}^{-q_{i}^{2} / 2}d\mu_{i}) =
d\omega \otimes W_{i} {\rm e}^{-q_{i}^{2} / 2}d\mu_{i} + (-)^{p} \omega
(W'_{i} - q_{i}W_{i})d\mu_{i}. \cr }
$$
Notice $W'= {\partial W \over \partial q_{i}}$ is an odd function of $q_{i}$.
We also set $\tilde d \, d\mu_{\phi}=0$.

The rationale for this is the following. Let $ \pi_* $ denote the
integration along the fibres of $\tilde {\cal N}$. We have
$$
\eqalign {
&d \, \pi_* (\omega \otimes W_{i}{\rm e}^{-q_{i}^{2} / 2}d\mu_{i})=0 \cr
&\pi_* \tilde d (\omega \otimes W_{i}{\rm e}^{-q_{i}^{2} / 2}d\mu_{i})=
\pi_*(d\omega \otimes W_{i}{\rm e}^{-q_{i}^{2} / 2}d\mu_{i} + (-)^{p}
\omega (W'_{i}- q_{i}W)d\mu_{i}) =0 \cr }
$$
and  $d \, \pi_*(\omega \otimes W d\mu_{\phi})= d(f\omega)=fd\omega +df \omega$
with $f=\int W d\mu_{\phi}$. But now
$$
\eqalign {
\pi_* \tilde d (\omega \otimes  W d\mu_{\phi})&=
\pi_* (d\omega \otimes Wd\mu_{\phi} + dW \wedge \omega d\mu_{\phi}) = \cr
&= d\omega f + df \, \omega \cr }
$$
and in all cases $d \, \pi_*= \pi_* \tilde d$.

The space $H^{p,0}(\tilde d) = {\rm Ker} \,  \tilde d / {\rm Im} \, \tilde d$
is then obviously isomorphic to $H^{p}({\cal M})$. Accordingly, we identify
$\tilde d$ with the functional counterpart of the BRS operator: up to
gaussian integrations it coincides with the exterior differential on the moduli
spaces.

Let us now go back to observables. Given cycles $c_i\; (i=1,2)$
representing homology classes $[c_i]\in H_*({\cal M},\Zz )$ of
complementary dimension, we can compute their intersection
$c_1.c_2\in\Zz$. We can easily construct an observable on $\tilde {\cal
N}$ whose expectation value equals $c_1.c_2$. Indeed we have
$$c_1.c_2=\int _{\cal M} \alpha_1\wedge \alpha_2,$$
where $[\alpha_i]\in H^*({\cal M},\Rr )$ is the Poincar\'e dual of
$[c_i]$. Setting $\alpha _1\wedge \alpha_2=fdv$, we simply multiply by
the "Thom class" $d\mu _\phi$ getting the measure $fd\mu$ on
$\tilde{\cal N}$. Accordingly, the expectation value of $f$ reads
$$<f>=:\int_{\tilde{\cal N}}fd\mu = \int_{\cal M} fdv=c_1.c_2.$$

Summing up, with the identification made above, we have seen that

{\bf Proposition.} There are
observables in the theory whose expectation values is the same as the
intersection of cycles in the moduli spaces.

\vfill\eject
\noindent {\bf 5. Concluding Remarks }

The basic result of this paper is that semiclassical field theories with
instantons moduli spaces have ''observables" corresponding to the intersection
theory on ${\cal M}$. As it is obvious, such expectation values do not depend
on the family $C_{\phi}$, $\phi \in {\cal M}$, of covariances given by the
field
theory itself and this has the following consequences:

\noindent
i) as for the expectation values of topological observables,
one can safely set $C_{\phi}=0$ and kill all quantum degrees of freedom in
the theory. This is the same as projecting $\tilde {\cal N}$ onto ${\cal M}$
and work out intersection theory there by topological methods. This is the
spirit of topological field theories.

\noindent
ii) the topological sector is by its very definition non perturbative: after
all
changing the measures and their support (e.g. by renormalization) does not
change intersection of cycles. So we see that keeping some non linearity in
$\tilde {\cal N}$ (represented by its base space ${\cal M}$) has the effect of
extracting non perturbative informations.

On the other hand we have seen that
quantum field theory and the topology of ${\cal M}$ do not mix up  at
least in our set up. Nor do we believe that this phenomenon arises from the
technical assumptions we made in building up our gaussian cohomology.
As far as some form of the Thom isomorphism holds in the infinite dimensional
case, the conclusion will be invariably the same. Although this makes hard to
immagine to compute intersection theory via quantum field theory, it makes
feasible to compute non perturbative effects via topological methods.

\vskip 2 truecm

\noindent {\bf Acknowledgements :} We thank E.Aldrovandi, L.Alvarez-Gaum\'e,
F.Cipriani, D. Franco, R. Kerner , E.Rabinovici and S.Vidussi for discussion.
We are also grateful to M.Carfora for suggesting the example 3.2.
C.R. acknowledges CERN and LPT in Paris for stages when part of this
work was developed.

\vskip2truecm
\noindent {\bf Appendix.}
\vskip3truemm
In the construction of the "gaussian cohomology" of section 4  we
assumed that there are global non vanishing sections of ${\cal N}'$. In
other words, we need a current on which the covariances $C_\phi$ does
not vanish for any $\phi\in {\cal M}$. This is the same as finding a section
$j$ of ${\cal N}$ such that $H_\phi j \not=0 \forall \phi \in {\cal M}$.
We will prove here that this is indeed the case for gauge theories
and Kahler manifold-valued $\sigma$-models.

\vskip3truemm \noindent A.1 -. As for gauge theories, recall that the
measure is carried on the family of dual spaces to the nuclear
completion ${\cal N}'\oplus {\cal I}'$ of the extension
of the conormal bundle ${\cal N}'_s$ by the bundle ${\cal I}'_{s-1}$ on
${\cal M}$ with fibre the dual of the Lie algebra of
the group of gauge transformations. This plays the
r\^ ole of the normal bundle in the present case.
Besides gauge fixing problems, this extension is technically usefull
for us because any
$0\not= \xi\in \Omega^0(adP)$ gives a section of the
(enlarged) normal bundle on the moduli spaces of irreducible
instantons along which the full covariance does not
vanish. In fact $ker\vartriangle ^2_A=ker\vartriangle _A=kerd_A=0\in
\Omega^0(adP)$ at any irreducible instanton $A$.

\vskip3truemm \noindent A.2 -. For Kahler manifold-valued
$\sigma$-models, the kernel of the Hessian operator at an instanton is
given by holomorphic vector fields along the instanton itself on the target
manifold, i.e. $kerH_\phi =H^0(C,\phi^*T'Y)$. The $C^\infty$ structure
of the vector bundle $\phi^*T'Y$ does not depend on the choice of
$\phi$ in a given connected component of the moduli instanton space.
Although the holomorphic structure of this bundle does depend on
$\phi$, any compactedly supported section with support contained in a
proper open subset of $C$ will never be holomorphic at any $\phi$, and
hence, using the fibre metric, will give us a current on which the
covariance never vanishes.

\vskip 2 truecm
\noindent {\bf References}

\noindent [A] M.F.Atiyah  {\it K--theory}   Benjamin, New York (1967)

\noindent [AB] M.F.Atiyah  R.Bott  Phil.Trans.R.Soc.Lond. A308, 523-615 (1982)

\noindent [Be] A.L. Besse {\it Einstein manifolds} Springer, New York
(1987)

\noindent [DK] S.K.Donaldson P.B.Kronheimer  {\it The geometry of
four--manifolds}  Clarendon Press--Oxford (1990)

\noindent [E] D.G. Ebin Proc. Symp. AMS n.15, Global Analysis p. 11
(1968)

\noindent [EW] J.Eells J.C. Wood Adv. Math. 49, 217 (1983)

\noindent [GV]  I.M.Gel'fand  N.Y.Vilenkin  {\it Generalized Functions}  vol.4
Academic Press (1964)

\noindent [G]   D.Gepner    Comm.Math.Phys. 141, 381 (1991)

\noindent [GP] D.Grossier  T.H.Parker  Comm.Math.Phys. 135, 101 (1990)

\noindent [L] J.M.F.Labastida  Comm.Math.Phys.  123, 641 (1989)

\noindent [O] H.Omori  {\it Infinite dimensional Lie transformations groups}
Springer (1974)

\noindent [P]  R.S.Palais  {\it Fundation of Global Analysis} Benjamin,
New York (1968)

\noindent [S] I.M.Singer  Comm.Math.Phys.  60, 7 (1978)

\noindent [St] S.A. Stromme "On parametrized rational curves in
Grassmann varieties" in {\it Space Curves}, F. Ghione et al ed.s - Lecture
Notes in Mathematics - Springer (1987)

\noindent [SY] Y.-T.Siu S.-T.Yau  Inventiones math. 59,189 (1980)

\noindent [`tH] G.`t Hooft  Phys. Rev. Lett.  37 n.1 (1976)

\noindent [W] E.Witten  Comm.Math.Phys.  117, 353 (1988)

\noindent [Y] Y.Yamasaki {\it Measures on Infinite Dimensional Spaces}
World Scientific (1985)

\vfill
\eject

\bye